\documentclass[pdflatex,sn-mathphys-num]{sn-jnl}
\usepackage{graphicx}
\usepackage{multirow}%
\usepackage{amsmath,amssymb,amsfonts}%
\usepackage{amsthm}%
\usepackage{mathrsfs}%
\usepackage[title]{appendix}%
\usepackage{xcolor}%
\usepackage{textcomp}%
\usepackage{manyfoot}%
\usepackage{booktabs}%
\usepackage{algorithm}%
\usepackage{algorithmicx}%
\usepackage{algpseudocode}%
\usepackage{listings}%
\usepackage{float}%
\usepackage{tabularx}%
\usepackage{array}%
\usepackage{capt-of}%
\usepackage{physics}%
\usepackage{mathtools}%
\usepackage{ragged2e}%
\usepackage{bm}%

\newcolumntype{L}{>{\RaggedRight\arraybackslash}X}

\theoremstyle{thmstyleone}%
\theoremstyle{thmstyletwo}%

\theoremstyle{thmstylethree}%

\begin{document}

\title[CV-QKD with VPS and QMSD]{CV-QKD over Turbulence Channels with Virtual Photon Subtraction and Quantum Multiple-Symbol Detection for Underwater Quantum Communications}


\author*[1]{\fnm{Nour} \sur{Rizk}}
\email{nour.rizk@ensta.fr}

\author[1]{\fnm{Hesham S.} \sur{Ibrahim}}

\author*[1]{\fnm{Ang\'elique} \sur{Dr\'emeau}}
\email{angelique.dremeau@ensta.fr}

\author*[1]{\fnm{Arnaud} \sur{Coatanhay}}
\email{arnaud.coatanhay@ensta.fr}

\affil[1]{\orgdiv{Lab-STICC, UMR CNRS 6285}, \orgname{ENSTA, Institut Polytechnique de Paris}, \orgaddress{\city{Brest Cedex 9}, \postcode{29200}, \country{France}}}

\abstract{Continuous-variable quantum key distribution (CV-QKD) is a promising approach for secure underwater quantum communications (UQCs), where propagation loss, scattering, turbulence, and receiver thermal noise can severely degrade the transmission of quantum states. In this paper, we propose an underwater CV-QKD system with virtual photon subtraction (VPS), implemented through post-selection of Alice’s measurement outcomes, without requiring channel state information (CSI) at the receiver. Three VPS-based system configurations are analyzed, corresponding to homodyne detection (VPS-HD), quantum maximum-likelihood detection (VPS-QMLD), and quantum multiple-symbol detection (VPS-QMSD). System performance is evaluated in terms of the accepted-only quantum bit error rate (QBER), where underwater turbulence is modeled by an Erlang distribution. Analytical and semi-closed-form QBER expressions are derived for the three configurations and validated through Monte Carlo simulations for different water types and system parameters. The results show close agreement between analytical derivations and simulation outcomes and demonstrate that VPS-QMSD provides the best robustness against underwater turbulence, achieving the lowest QBER compared with VPS-QMLD and VPS-HD.}

\keywords{Continuous-variable quantum key distribution (CV-QKD), underwater quantum communications (UQCs), virtual photon subtraction (VPS), quantum maximum-likelihood detection (QMLD), quantum multiple-symbol detection (QMSD), quantum bit error rate (QBER)}

\maketitle

\section{Introduction}
Underwater wireless optical communication (UWOC) provides a high-bandwidth and low-latency alternative to acoustic links for critical marine applications, including ocean monitoring, offshore inspection, and autonomous underwater vehicle (AUV) coordination \cite{kaushal2016underwater,saeed2019underwater}. As these networks increasingly support security-sensitive missions, such as defense, surveillance,
and critical-infrastructure monitoring, confidentiality must be addressed together with transmission reliability \cite{aman2023security}. Conventional cryptographic techniques can protect classical communication links, but their security ultimately relies on computational assumptions. Quantum key distribution (QKD), by contrast, enables two legitimate users to establish shared secret keys with security based on quantum-mechanical principles, provided that physical implementation and post-processing assumptions are properly considered \cite{scarani2009security,Pirandola2020}.

Early studies on underwater QKD (UQKD) have mainly focused on discrete-variable QKD (DV-QKD) protocols, including prepare-and-measure schemes such as BB84, SARG04, and decoy-state protocols, as well as entanglement-based approaches such as E91 and BBM92  \cite{rizk2025performances,rizk2026bbm92,rizk2026performance,rizk2026nonmaximally,rizk2026eusipco,brassard1984quantum,scarani2004quantum,hwang2003quantum,lo2005decoy,ekert1991quantum,bennett1992quantum}. In these protocols, information is encoded into discrete quantum states, such as polarization states, or extracted from entanglement correlations. However, underwater propagation imposes severe physical impairments (such as absorption, depolarizing effects, scattering, and turbulence-induced fading) alongside high background irradiance. These factors drastically reduce the received photon flux and increase the quantum bit error rate (QBER) ~\cite{mobley1994light,zhao2019performance,yang2026impact}. In addition, practical DV-QKD systems often employ attenuated coherent laser pulses rather than ideal single-photon sources. Without appropriate countermeasures, the multi-photon component of weak coherent pulses can be exploited by photon-number-splitting (PNS) attacks; decoy-state protocols mitigate this vulnerability but do not remove the underwater-channel penalties associated with loss, scattering, alignment, detector dead time, and background noise \cite{brassard2000limitations,hwang2003quantum,lo2005decoy,dong2022practical}. These factors make the practical deployment of underwater DV-QKD highly challenging.

These limitations motivate the transition toward continuous-variable QKD (CV-QKD), which encodes information in field quadratures of coherent or squeezed optical states and recovers it using coherent detection \cite{grosshans2002continuous, grosshans2003quantum, diamanti2015distributing}. Compared with DV-QKD, CV-QKD is attractive for short- and medium-range optical links because it can exploit standard optical communication components, high-efficiency coherent receivers, and mature digital signal processing. The experimental feasibility of underwater CV-QKD has been demonstrated by Tang et al., who implemented a discrete-modulated CV-QKD system over an underwater channel and reported a secure key rate of 22.9 kbits/s at a channel loss of 12.4 dB, corresponding to an equivalent distance of 148.7 m in pure seawater \cite{tang2022experimental}.

Nevertheless, the practical advantages of CV-QKD do not fully overcome the impairments imposed by underwater propagation. Absorption and scattering reduce the received optical power, while turbulence introduces random fading that affects the detection reliability. To improve robustness under loss and noise, non-Gaussian operations have been investigated in CV-QKD \cite{hu2020continuous,navarrete2012enhancing}. Among these operations, photon subtraction can increase effective entanglement and improve tolerance to channel noise. Yet, its direct physical implementation requires conditional photon detection and therefore suffers from limited success probability and additional experimental complexity. Virtual photon subtraction (VPS) provides a practical alternative by reproducing the effect of photon subtraction through post-selection of Alice’s measurement outcomes \cite{li2016nongaussian}. Recent oceanic CV-QKD studies have considered VPS combined with homodyne detection and evaluated channel-dependent performance using ABCD-matrix-based underwater propagation models \cite{meena2025continuous, kogelnik1965imaging, kogelnik1965propagation}. 

However, improving the source alone is not sufficient, since the receiver must also make reliable decisions from observations degraded by underwater channel impairments. In this context, homodyne detection (HD) serves as a natural coherent-detection benchmark, whereas photon-counting reception with photon-number-resolving (PNR) detectors provides discrete photon-count observations suitable for quantum maximum-likelihood detection (QMLD). Maximum-likelihood approaches have been widely investigated in quantum-state estimation and detection, starting from quantum-state maximum-likelihood schemes based on positive operator-valued measure (POVM) outcomes, and later extended to the reconstruction of mixed quantum states from noisy measurement outcomes, including additive Gaussian noise \cite{hradil1997quantum, smolin2012efficient}. In standard QMLD approaches, each transmitted bit or symbol is inferred from its corresponding received observation. Such symbol-by-symbol processing does not exploit the fact that underwater turbulence may remain correlated over several consecutive channel uses. The quantum multiple-symbol detection (QMSD) scheme addresses this limitation by jointly processing a block of received observations and marginalizing the unknown fading state, enabling detection without requiring channel-state information (CSI-free detection) under block-correlated turbulence \cite{dong2025coherent}.

Although VPS-based source engineering and QMSD-based receiver processing have been studied separately, their combination has not been adequately investigated for underwater CV-QKD. Existing QMSD studies have mainly considered coherent-state transmission over turbulent channels, without accounting for the accepted non-Gaussian source distribution induced by VPS. Conversely, existing VPS-based underwater CV-QKD studies have primarily relied on homodyne-detection-based performance analysis and have not considered photon-counting QMLD/QMSD receivers. The joint effect of VPS post-selection, underwater fading, PNR photon-counting statistics, and CSI-free block detection therefore remains an open performance question.

In this paper, we investigate a CSI-free underwater CV-QKD system based on virtual photon subtraction at Alice’s side and photon-counting or homodyne reception at Bob's side. The proposed framework combines the accepted non-Gaussian source induced by VPS with three receiver configurations: VPS-HD, VPS-QMLD, and VPS-QMSD. The main contributions of this work are summarized as follows:
\begin{itemize}
    \item We develop a system model of the considered VPS-based underwater CV-QKD scheme. Starting from the entanglement-based photon-subtraction description, we derive the equivalent prepare-and-measure transmitter model and characterize the accepted non-Gaussian source distribution generated by VPS post-selection.
    \item We derive the photon-counting statistics at the receiver after underwater propagation, displacement operations, and PNR detection. In particular, we obtain the bit-conditioned photon-count likelihood and a semi-closed-form CSI-free expression under deterministic path loss, Erlang-modeled turbulence, and receiver thermal noise.
    \item We formulate an analytical framework for evaluating the accepted-only QBER of the three receiver schemes. The resulting expressions provide semi-closed-form performance characterizations under underwater channel impairments and are validated by Monte Carlo simulations for different water types and system parameters.
\end{itemize}

The remainder of this paper is organized as follows. Section \ref{section2} presents the proposed VPS-based underwater CV-QKD system model, including the source description, underwater channel model, and receiver statistics. Section \ref{QBER} develops the detection rules and accepted-only QBER expressions for the VPS-QMLD, VPS-QMSD, and VPS-HD schemes. Section \ref{Results} presents the numerical results and Monte Carlo validation. Finally, Section \ref{Conclusion} concludes the paper.

\section{VPS-Based Transmitter Model}
\label{section2}
This section presents the system model of the considered CV-QKD scheme employing photon subtraction, as illustrated in Fig. \ref{fig_system_model}. We begin by reviewing the principle of physical photon subtraction applied to a two-mode squeezed vacuum (TMSV) state and then introduce its equivalent virtual implementation via post-selection. Next, we develop the prepare-and-measure transmitter model corresponding to the entanglement-based description and characterize the accepted non-Gaussian source distribution induced by virtual photon subtraction. Finally, we present the channel model and the receiver structure used for detection and performance analysis.

\begin{center}
\includegraphics[width=1.0\textwidth]{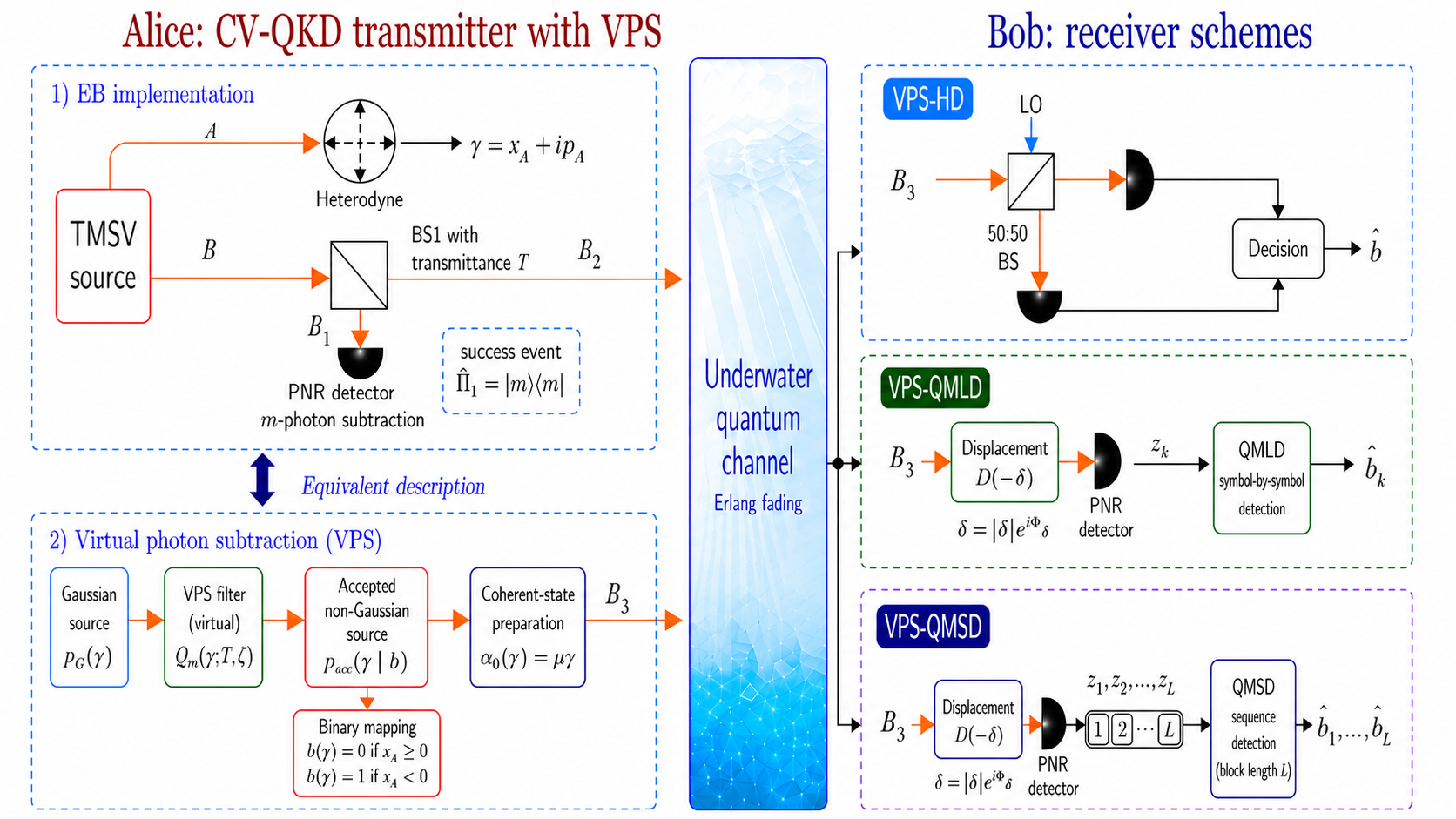}
\captionof{figure}{Schematic representation of the CV-QKD scheme with VPS over an underwater quantum channel (UQCs). Alice's transmitter is shown through the EB photon-subtraction implementation and its equivalent prepare-and-measure VPS description.
The received mode \(B_3\) is processed at Bob using one of three receiver schemes: VPS-HD, VPS-QMLD, or VPS-QMSD. $BS1$: beam splitter; \mbox{$\gamma$: Alice's measurement result;} $\zeta$: TMSV parameter; $Q_m(\gamma;T,\zeta)$: postselection filter function.}
\label{fig_system_model}
\end{center}

\subsection{Photon Subtraction}

Photon subtraction is a non-Gaussian operation that enhances the entanglement of a two-mode squeezed vacuum (TMSV) state. Several photon subtraction strategies have been investigated in the literature, including operations applied to either a single mode or both modes of the TMSV state \cite{li2016nongaussian}. In this subsection, we first review the principle of physical photon subtraction and then introduce an equivalent non-Gaussian post-selection strategy for CV-QKD, referred to as virtual photon subtraction (VPS).

\subsubsection{Physical Photon Subtraction}

The entanglement-based (EB) implementation of the CV-QKD protocol incorporating photon subtraction at Alice’s station is illustrated in Fig. \ref{fig_system_model}. After generating the TMSV state, which consists of two modes $A$ and $B$, Alice applies a beam splitter (BS1) with transmittance $T$ to split mode $B$ into two output modes, $B_1$ and $B_2$. Subsequently, mode $B_1$ is measured using a positive operator-valued measure (POVM) $\{\hat\Pi_0,\hat\Pi_1\}$. The photon subtraction operation is considered successful when the measurement yields the outcome associated with $\hat\Pi_1$. Conditioned on this event, the remaining modes $A$ and $B_2$ are retained, resulting in the photon-subtracted TMSV state.

More generally, the subtraction operation can be extended to the removal of $m$ photons. In this case, the corresponding POVM element is $\hat{\Pi}_1 = |m\rangle\langle m|$, which can be implemented using a PNR detector \cite{meena2025continuous}.
Previous studies have demonstrated that the entanglement of the resulting state increases with the number of subtracted photons \cite{navarrete2012enhancing}.

\subsubsection{Virtual Photon Subtraction via Post-Selection}

Consider a CV-QKD system in which Alice employs a photon-subtracted TMSV state as the entanglement source and performs heterodyne detection on mode $A$. 
In the entanglement-based scheme, Alice records whether the subtraction condition associated with the POVM element $\hat{\Pi}_1$ is satisfied for each generated state and uses this information to decide whether the corresponding realization is retained or discarded. For a PNR detector, $\hat{\Pi}_1$ may represent the subtraction of a specific photon number, whereas for an on-off detector, this reduces to a click/no-click decision. After Bob performs his measurement on the received mode, Alice announces the acceptance decisions during the classical post-processing.


An equivalent implementation can be achieved through virtual photon subtraction employing a post-selection strategy based on Alice’s measurement outcomes. Instead of physically implementing the photon subtraction operation, Alice performs standard measurements and subsequently applies a probabilistic filtering procedure to her data that reproduces the statistical effect of photon subtraction. This virtual implementation provides two main advantages. First, it eliminates the need for dedicated physical hardware, thereby reducing the system's experimental complexity. Second, it can achieve improved performance compared with its physical counterpart, as the post-selection procedure can effectively emulate the behavior of an ideal photon-number detector.

Since the heterodyne measurement on mode $A$ and the POVM measurement on mode $B_1$ act on different modes, they commute. Consequently, Alice may first perform heterodyne detection on mode $A$, followed by a POVM measurement on mode $B_1$. It is well established that heterodyne detection applied to one mode of the TMSV state results in the projection of the complementary mode onto a coherent state. This observation provides the basis for the prepare-and-measure description of virtual photon subtraction developed next.

\subsection{Prepare-and-Measure Equivalent Source}

We adopt a prepare-and-measure (PM) description that is equivalent to an entanglement-based implementation where Alice performs heterodyne detection and, conditioned on her measurement outcome, prepares a coherent state for transmission. Accordingly, for a heterodyne outcome $\gamma = x_A + i p_A$, the state of modes $B_1$ and $B_2$ after passing through BS1 can be written as \cite{meena2025continuous}
\begin{equation}
|\psi\rangle_{B_1B_2} =
\left|\sqrt{1-T}\,\alpha \right\rangle_{B_1}
\left|\sqrt{T}\,\alpha \right\rangle_{B_2},
\end{equation}
where $T$ denotes the transmittance of BS1, $\alpha = \frac{\sqrt{2}\zeta}{2}\gamma$, and $\zeta$ representing the TMSV parameter. Equivalently, the coherent state prepared for transmission in the PM description has an amplitude
\begin{equation}
\alpha_o(\gamma)=\sqrt{T}\,\alpha(\gamma)
=\mu\gamma,
\qquad
\mu =\frac{\sqrt{2T}\,\zeta}{2}.
\end{equation}
The variable $\gamma$ follows a complex Gaussian distribution given by
\begin{equation}
p_G(\gamma)=\frac{1}{\pi (V_T + 1)}\exp\!\left(-\frac{|\gamma|^2}{V_T + 1}\right),
\label{eq:pG}
\end{equation}
where $V_T = \frac{(1+\zeta^2)}{(1-\zeta^2)}$ is the variance of the TMSV state.

For the subtraction of $m$ photons, the conditional success probability associated with the measurement outcome $\gamma$ is given by \cite{meena2025continuous}
\begin{equation}
Q_m(\gamma,T,\zeta)
=
p^{\hat{\Pi}_1}(m|x_A,p_A)
=
\exp\!\left(-y|\gamma|^2\right)
\frac{\left(y|\gamma|^2\right)^m}{m!},
\qquad
y=\frac{(1-T)\zeta^2}{2}.
\label{eq:Qm}
\end{equation}

This quantity acts as the post-selection filter function in the virtual protocol. Each data point is accepted with probability $Q_m(\gamma,T,\zeta)$, thereby reproducing the effect of physical photon subtraction at the level of the retained statistics.

The overall acceptance probability, i.e., the fraction of transmitted attempts that survive the VPS filter, is obtained by averaging the filter function over the Gaussian prior distribution of $\gamma$:
\begin{equation}
P_{\mathrm{acc}}
=
\int_{\mathbb{C}} Q_m(\gamma,T,\zeta)\,p_G(\gamma)\,d^2\gamma.
\label{eq:Pacc1}
\end{equation}
Substituting Eqs.~\eqref{eq:pG} and \eqref{eq:Qm} into Eq.~\eqref{eq:Pacc1}, we obtain
\begin{equation}
P_{\mathrm{acc}}
=
\frac{y^m}{m!\,\pi (V_T+1)}
\int_{\mathbb{C}}
|\gamma|^{2m} e^{-w|\gamma|^2}\, d^2\gamma,
\qquad
w = y+\frac{1}{V_T+1}.
\label{eq:Pacc2}
\end{equation}
Evaluating the integral in polar coordinates, where $\gamma = r e^{i\phi}$ and $d^2\gamma = r\,dr\,d\phi$, yields
\begin{equation}
P_{\mathrm{acc}}
=
\frac{y^m}{(V_T+1)\,w^{m+1}}.
\label{eq:Pacc_final}
\end{equation}

Conditioned on acceptance, the retained measurement outcomes follow the distribution
\begin{equation}
p_{\mathrm{acc}}(\gamma)
=
\frac{Q_m(\gamma,T,\zeta)\,p_G(\gamma)}
{P_{\mathrm{acc}}}.
\label{eq:pacc_uncond}
\end{equation}
Therefore, VPS does not alter the initial Gaussian modulation itself, but probabilistically reshapes it into a non-Gaussian ensemble through post-selection.

Finally, in the PM-equivalent implementation, the effect of BS1 with transmittance $T$ is reproduced by generating a coherent state with the reduced mean amplitude $\sqrt{T}\alpha$. Hence, virtual photon subtraction is realized through two combined operations: post-selection of Alice's heterodyne data according to the filter function $Q_m(\gamma,T,\zeta)$, and rescaling of the coherent-state amplitude by the factor $\sqrt{T}$.


To interface the VPS source model with the subsequent quantum multiple-symbol detection (QMSD), we assign a binary label to each accepted heterodyne outcome according to the sign of Alice's in-phase quadrature,
\begin{equation}
b(\gamma)=
\begin{cases}
0, & x_A \ge 0,\\
1, & x_A < 0.
\end{cases}
\label{eq:binary_mapping}
\end{equation}
Using the indicator function $\mathbf{1}[\cdot]$, the accepted-only bit-conditioned density can be written as
\begin{equation}
p_{\mathrm{acc}}(\gamma \mid b)
=
\frac{
Q_m(\gamma,T,\zeta)\,p_G(\gamma)\,
\mathbf{1}\!\left[\operatorname{sign}(x_A)=(-1)^b\right]
}{
P_{\mathrm{acc}}^{(b)}
},
\label{eq:pacc_bit}
\end{equation}
where the per-bit acceptance probability is given by
\begin{equation}
P_{\mathrm{acc}}^{(b)}
=
\int_{\mathbb{C}}
Q_m(\gamma,T,\zeta)\,p_G(\gamma)\,
\mathbf{1}\!\left[\operatorname{sign}(x_A)=(-1)^b\right]
\, d^2\gamma.
\label{eq:Pacc_b}
\end{equation}
Since both the Gaussian prior $p_G(\gamma)$ and the VPS filter $Q_m(\gamma,T,\zeta)$ depend only on $|\gamma|^2$, the two bit regions have equal probability mass. Hence,
\begin{equation}
P_{\mathrm{acc}}^{(0)}=P_{\mathrm{acc}}^{(1)}=\frac{P_{\mathrm{acc}}}{2}.
\label{eq:Pacc_bit_symmetry}
\end{equation}
This binary mapping provides the interface between the accepted non-Gaussian VPS source distribution and the subsequent QMSD receiver model.

\section{Underwater Quantum Channel Model and Receiver Statistics}
\label{Channel_model}
The propagation of quantum states through an underwater quantum channel is subject to average attenuation due to absorption and scattering, as well as random intensity fluctuations caused by marine turbulence. Both effects alter the received optical field and must therefore be incorporated into the channel model used for detection and performance analysis.

\subsection{Underwater Channel Model }

As introduced above, the deterministic attenuation of optical signals in underwater environments is governed by absorption and scattering phenomena. These effects are jointly characterized by the extinction coefficient $c(\lambda)$, defined as \cite{mobley1994light}
\begin{equation}
c(\lambda)=a(\lambda)+b(\lambda),
\label{eq:extinction_coeff}
\end{equation}
where $a(\lambda)$ and $b(\lambda)$ are the absorption and scattering coefficients at wavelength $\lambda$, respectively. For coherent states propagating over a distance $d$, the path-loss coefficient follows the Beer--Lambert law as
\begin{equation}
I_p=e^{-c(\lambda)d}.
\label{eq:Ip}
\end{equation}

\label{turbulance_channel}
In addition to the path loss, marine turbulence induces random irradiance fluctuations at the receiver and thus significantly affects underwater quantum communication performance. The normalized intensity fluctuation, denoted by $I_t$, characterizes the random variations in the amplitude of the received coherent states. Under weak-to-moderate turbulence conditions, $I_t$ is modeled by a log-normal distribution whose probability density function (PDF) is given by \cite{kaushal2016underwater, chen2023optimal}
\begin{equation}
p_{\mathrm{L}}(I_t)
=
\frac{1}{I_t\sqrt{2\pi\sigma_X^2}}
\exp\!\left[
-\frac{1}{2\sigma_X^2}
\left(\ln I_t + \frac{\sigma_X^2}{2}\right)^2
\right],
\qquad I_t>0,
\label{eq:lognormal_pdf}
\end{equation}
where $\sigma^2_X$ is the log-amplitude variance, which characterizes the turbulence strength.

While the log-normal model is physically pertinent, its direct application in the likelihood analysis results in analytically cumbersome expressions. To reduce complexity and facilitate the evaluation of the resulting integrals, the log-normal distribution is approximated by an Erlang distribution via mean and variance matching, while preserving the positive-valued support required for intensity fading \cite{abourjeily2010achievable, riediger2009fast}. The resulting PDF is characterized by the shape parameter $\theta\in\mathbb{Z}^{+}$ and the rate parameter $\lambda_E>0$ \cite{papoulis1991probability}:
\begin{equation}
p_{\mathrm{E}}(I_t)
=
\frac{\lambda_E^\theta}{(\theta-1)!}\,
I_t^{\theta-1}e^{-\lambda_E I_t},
\qquad I_t>0,
\label{eq:erlang_pdf}
\end{equation}

Combining the path-loss term and the turbulence term, the overall channel gain between the transmitter and the receiver is expressed as \cite{rahman2024performance}
\begin{equation}
I = I_p I_t.
\label{eq:I_total}
\end{equation}
Consequently, the received coherent-state amplitude before any further processing becomes:
\begin{equation}
\alpha_{\mathrm{or}}(\gamma,I_t)=\sqrt{I_p I_t}\,\mu\gamma.
\label{eq:alphar}
\end{equation}

\subsection{Receiver Model and Photon-Counting Statistics}

At the receiver side, a displacement-based PNR detector is adopted to enhance the binary discrimination of the VPS-encoded signal after underwater propagation. In VPS modulation, each transmitted bit maps to a non-Gaussian ensemble of coherent states rather than a single phase-space point. The underwater channel further distorts these states via path loss, turbulence-induced fading, and background noise, causing significant overlap between the conditional photon-count distributions of the two bit classes, thereby degrading direct detection performance. To address this, a coherent displacement operation is applied to the received state to optimally separate the two distributions in phase space, after which the PNR detector performs a binary decision based on the resulting count statistics.
The receiver displacement is represented by the complex parameter
\begin{equation}
\delta=|\delta|e^{i\phi_\delta}.
\label{eq:delta}
\end{equation}
After the application of the displacement operator $D(-\delta)$, the coherent amplitude at the detector input becomes
\begin{equation}
\alpha_{\mathrm{ord}}(\gamma,I_t,\delta)=\sqrt{I_pI_t}\,\mu\gamma-\delta,
\label{eq:xi}
\end{equation}
and the corresponding displaced energy is
\begin{equation}
S_\delta(\gamma,I_t)
=\left|\alpha_{\mathrm{ord}}(\gamma,I_t,\delta)\right|^2 \\
=\left|\sqrt{I_pI_t}\,\mu\gamma-\delta\right|^2, \\
\end{equation}
which can be expanded as
\begin{equation}
    S_\delta(\gamma,I_t)= I_p I_t \mu^2|\gamma|^2 + |\delta|^2
-2\Re\!\left\{\delta^* \sqrt{I_pI_t}\,\mu\gamma\right\},
\label{eq:Sdelta}
\end{equation}
where $\mu \in \mathbb{R}_{+}^{*}$.

Assuming thermal noise with a mean photon number $N>0$, the conditional probability of observing $z$ photons at the PNR detector is given by the displaced thermal-state count law
\begin{equation}
\Pr(z\mid \gamma,I_t,\delta)
=
\frac{N^z}{(N+1)^{z+1}}
\exp\!\left(-\frac{S_\delta(\gamma,I_t)}{N+1}\right)
L_z\!\left(-\frac{S_\delta(\gamma,I_t)}{N(N+1)}\right),
\qquad z=0,1,2,\dots
\label{eq:PNR}
\end{equation}
where $L_z(\cdot)$ denotes the Laguerre polynomial of order $z$. The noiseless case $N=0$ is not obtained by direct substitution in Eq.~\eqref{eq:PNR}; it is recovered through the Poisson limit, $\Pr(z\mid \gamma,I_t,\delta)=e^{-S_\delta}S_\delta^z/z!$. For subsequent analytical developments, the Laguerre polynomial in Eq.~\eqref{eq:PNR} is expanded as
\begin{equation}
L_z(-u)=\sum_{k=0}^{z}\binom{z}{k}\frac{u^k}{k!},
\qquad u\ge 0.
\label{eq:LaguerreExpand}
\end{equation}

\subsection{Bit-Conditioned Photon-Count Likelihood}
\label{Bit-Conditioned Photon-Count Likelihood}

Based on the conditional PNR count law in Eq. \eqref{eq:PNR}, the photon-count likelihood for binary detection is obtained by averaging over the accepted VPS source distribution associated with each bit class. Conditioned on $I_t$, it takes the form
\begin{equation}
\Pr(z\mid b,I_t,\delta)
=
\int_{\mathbb{C}}
\Pr(z\mid \gamma,I_t,\delta)\,
p_{\mathrm{acc}}(\gamma\mid b)\,d^2\gamma.
\label{eq:PrzbItd_def}
\end{equation}
Substituting Eq. \eqref{eq:pacc_bit} into Eq. \eqref{eq:PrzbItd_def}, we obtain
\begin{equation}
\Pr(z\mid b,I_t,\delta)
=
\frac{1}{P_{\mathrm{acc}}^{(b)}}
\int_{\mathbb{C}}
\Pr(z\mid \gamma,I_t,\delta)\,
Q_m(\gamma,T,\zeta)\,p_G(\gamma)\,
\mathbf{1}\!\left[\operatorname{sign}(x_A)=(-1)^b\right]
\,d^2\gamma.
\label{eq:PrzbItd_expanded}
\end{equation}

Expressing $\gamma$ in polar coordinates, exploiting the binary partition induced by the sign of $x_A$, and using Eq.~\eqref{eq:Pacc_bit_symmetry}, Eq. \eqref{eq:PrzbItd_expanded} can be reformulated as
{\small
\begin{align}
\Pr(z\mid b,I_t,\delta)
=
\frac{2y^m N^z}{m!\,\pi(V_T+1)\,P_{\mathrm{acc}}\,(N+1)^{z+1}}
&\int_0^\infty r^{2m+1}e^{-wr^2}\notag\\
&\times\left[
\int_{\Phi_b}
e^{-S_\delta/(N+1)}
L_z\!\left(-\frac{S_\delta}{N(N+1)}\right)
\,d\phi
\right]dr,
\label{eq:PrzbItd_angular_exact}
\end{align}}
with
\begin{equation}
\left\{
\begin{aligned}
&w = y + \frac{1}{V_T+1},
\qquad
\Phi_0=\left[-\frac{\pi}{2},\frac{\pi}{2}\right),
\qquad
\Phi_1=\left[\frac{\pi}{2},\frac{3\pi}{2}\right), \\[0.8em]
&S_\delta(r,\phi,I_t)
=
I_p I_t \mu^2 r^2 + |\delta|^2
-2|\delta|\,\mu\sqrt{I_p I_t}\,r\cos(\phi-\phi_\delta).
\end{aligned}
\right.
\label{eq:wPhiSdelta}
\end{equation}

The detailed derivation of Eq. \eqref{eq:PrzbItd_angular_exact}, as well as the resulting semi-closed-form representation of $\Pr(z\mid b,I_t,\delta)$, is provided in Appendix~\ref{Appendix_A}.

\subsection{Semi-closed-form CSI-free likelihood}
\label{Semi-closed-form CSI-free likelihood}
The likelihood derived in Subsection~\ref{Bit-Conditioned Photon-Count Likelihood} remains conditioned on the turbulence term $I_t$. When this channel information is not available at the receiver, the detection rule must instead rely on a likelihood averaged over the channel statistics. Using the Erlang distribution introduced in Subsection~\ref{turbulance_channel}, the CSI-free likelihood is expressed as
\begin{equation}
\Pr(z\mid b,\delta)
=
\int_0^\infty
\Pr(z\mid b,I_t,\delta)\,
p_{\mathrm{E}}(I_t)\,dI_t.
\label{eq:Przbd}
\end{equation}
Although this expression defines the relevant CSI-free likelihood, its direct evaluation remains analytically cumbersome. In particular, the conditional likelihood $\Pr(z\mid b,I_t,\delta)$ involves a coupled combination of Laguerre polynomials, exponential terms, and polynomial factors, which must subsequently be averaged over the fading statistics. This makes the resulting likelihood inconvenient to manipulate in the developments presented in the following sections. To address this issue, a semi-closed-form representation is derived to obtain a more manageable expression while preserving the dependence of the likelihood on the system and channel parameters. In the semi-closed-form expressions below, the displacement phase is chosen along the in-phase quadrature, i.e., $\phi_\delta=0$; the exact integral form in Eq.~\eqref{eq:PrzbItd_angular_exact} remains valid for arbitrary $\phi_\delta$. By substituting the semi-closed-form expression of $\Pr(z\mid b,I_t,\delta)$ derived in Appendix~\ref{Appendix_A} into Eq.~\eqref{eq:Przbd}, and then averaging with respect to the Erlang density, the CSI-free likelihood can be written as
\begin{equation}
\begin{aligned}
\Pr(z\mid b,\delta)
=
&\,
\mathcal{K}_z
\sum_{k=0}^{z}\binom{z}{k}\frac{1}{k!\,[N(N+1)]^k}
\sum_{n=0}^{k}\binom{k}{n}(-1)^n\Omega^n \\
&\times
\sum_{i=0}^{k-n}\binom{k-n}{i}
(|\delta|^2)^{k-n-i}(I_p\mu^2)^i
\sum_{j=0}^{\infty}\frac{\rho^j}{j!}\,
S_b(n,j)\,M_{n+j}\,
\mathcal{I}_{k,n,i,j},
\end{aligned}
\label{eq:Przbd_semiclosed}
\end{equation}
where $\{S_b(n,j)\}$ and $\{M_{n+j}\}$ denote the constants resulting from the angular integration over the bit-dependent domains. Using Eq.~\eqref{eq:Pacc_bit_symmetry}, the normalization factor is independent of $b$ and
\begin{equation}
\left\{
\begin{aligned}
\mathcal{K}_z
&=
\frac{N^z}{(N+1)^{z+1}}
\cdot
\frac{2y^m}{m!\,\pi(V_T+1)\,P_{\mathrm{acc}}}
\cdot
\frac{\lambda_E^\theta}{(\theta-1)!}
\cdot
\exp\!\left(-\frac{|\delta|^2}{N+1}\right),\\
\Omega&=2|\delta|\mu\sqrt{I_p},\qquad
\rho=\frac{\Omega}{N+1},\qquad
\chi=\frac{I_p\mu^2}{N+1},\\
\mathcal{I}_{k,n,i,j}
&=
\int_0^\infty
r^{2m+1+2i+n+j}
e^{-wr^2}
\frac{
\Gamma\!\left(\theta+i+\frac{n+j}{2}\right)
}{
(\lambda_E+\chi r^2)^{\theta+i+\frac{n+j}{2}}
}
\,dr.
\end{aligned}
\right.
\label{eq:Kz_new}
\end{equation}
The detailed derivation of Eq. \eqref{eq:Przbd_semiclosed} is presented in Appendix~\ref{Appendix_B}.

\section{Detection and Performance Analysis}
\label{QBER}
The likelihood expressions derived in Subsections~\ref{Bit-Conditioned Photon-Count Likelihood} and \ref{Semi-closed-form CSI-free likelihood} provide the basis for the detection and performance metrics developed in this section. In the considered VPS-assisted underwater quantum communication system, binary detection may be performed either independently for each observed photon count or jointly over a sequence of observations. To assess the benefit of sequence processing, two displaced photon-counting detectors are considered, namely quantum maximum-likelihood detection (QMLD) and quantum multiple-symbol detection (QMSD). In addition, homodyne detection (HD) is adopted as a benchmark receiver. Since virtual photon subtraction retains only a subset of the transmitted states, all performance measures reported in this section are conditioned on the acceptance event.

\subsection{Quantum Maximum-Likelihood Detection (QMLD)}
\label{subsec_QMLD}

For QMLD, each observed photon count is assigned to the bit hypothesis with the larger CSI-free likelihood, i.e.,
\begin{equation}
\hat b_L(z)=\arg\max_{b\in\{0,1\}} \Pr(z\mid b,\delta).
\label{eq:QMLD_rule}
\end{equation}

Assuming equiprobable binary hypotheses, the accepted-only QBER of QMLD is expressed as
\begin{equation}
\mathrm{QBER}_{\mathrm{QMLD,acc}}
=
\frac12
\sum_{z=0}^{\infty}
\min\left\{
\Pr(z\mid 0,\delta),\,
\Pr(z\mid 1,\delta)
\right\}.
\label{eq:QMLD_QBER}
\end{equation}

Equivalently, we define the maximum-likelihood decision regions as
\begin{equation}
\mathcal Z_0=
\left\{
z:\Pr(z\mid 0,\delta)\ge \Pr(z\mid 1,\delta)
\right\},
\qquad
\mathcal Z_1=
\left\{
z:\Pr(z\mid 1,\delta)>\Pr(z\mid 0,\delta)
\right\}.
\label{eq:QMLD_regions}
\end{equation}
Under this partition, Eq. \eqref{eq:QMLD_QBER} becomes
\begin{equation}
\mathrm{QBER}_{\mathrm{QMLD,acc}}
=
\frac12\sum_{z\in\mathcal Z_1}\Pr(z\mid 0,\delta)
+
\frac12\sum_{z\in\mathcal Z_0}\Pr(z\mid 1,\delta).
\label{eq:QMLD_QBER_regions}
\end{equation}

QMLD processes each received photon-count symbol independently, thereby neglecting the dependence that may arise across the observation sequence. This independent detection method may result in reduced detection performance, especially in the absence of CSI at the receiver.

\subsection{Quantum Multiple-Symbol Detection (QMSD)}
\label{subsec_QMSD}

To address the limitations of QMLD, QMSD is considered as a maximum-likelihood sequence detector in the absence of CSI at the receiver, jointly processing a block of $L$ received photon-count observations. Specifically, it identifies the most likely transmitted bit sequence from the full observation window of length $L$, as described below.

The received photon-count sequence over the observation window is denoted by
\begin{equation}
\mathbf z=(z_1,\dots,z_L),
\label{eq:QMSD_obs}
\end{equation}
while the corresponding transmitted bit sequence is given by
\begin{equation}
\mathbf b=(b_1,\dots,b_L)\in\{0,1\}^L.
\label{eq:QMSD_bits}
\end{equation}
Under the block-fading assumption, the corresponding CSI-free QMSD metric is therefore expressed as
\begin{equation}
\Lambda(\mathbf z\mid\mathbf b,\delta)
=
\int_0^\infty
\left[
\prod_{\ell=1}^{L}
\Pr(z_\ell\mid b_\ell,I_t,\delta)
\right]
p_{\mathrm E}(I_t)\,dI_t.
\label{eq:QMSD_metric}
\end{equation}
Based on this sequence metric, the detected bit sequence is obtained according to the maximum-likelihood criterion as
\begin{equation}
\hat{\mathbf b}_s(\mathbf z)
=
\arg\max_{\mathbf b\in\{0,1\}^L}
\Lambda(\mathbf z\mid\mathbf b,\delta).
\label{eq:QMSD_rule}
\end{equation}

Although Eq.~\eqref{eq:QMSD_metric} defines the QMSD metric in its general form, its direct evaluation remains analytically cumbersome, since it involves the fading average of a product of conditional symbol likelihoods. To facilitate the subsequent analysis, a semi-closed-form representation is derived under Erlang fading and is given by
\begin{equation}
\begin{aligned}
\Lambda(\mathbf z\mid \mathbf b,\delta)
=
\frac{\lambda_E^\theta}{\Gamma(\theta)}
\sum_{\kappa_1}\cdots\sum_{\kappa_L}
\left[
\prod_{\ell=1}^{L}
\Xi^{(b_\ell)}_{z_\ell;\kappa_\ell}
\right]
\chi^{-(\theta+\Sigma_s)}
w^{\theta+\Sigma_s-\Sigma_\tau}
\Gamma(\theta+\Sigma_s) \\
\times
U\!\left(
\theta+\Sigma_s,\,
\theta+\Sigma_s+1-\Sigma_\tau,\,
\frac{\lambda_E w}{\chi}
\right),
\end{aligned}
\label{eq:QMSD_metric_scf}
\end{equation}
where $\kappa_\ell=(k_\ell,n_\ell,i_\ell,j_\ell), \, 
\Sigma_s=\sum_{\ell=1}^{L}s_{\kappa_\ell},
\,
\Sigma_\tau=\sum_{\ell=1}^{L}\tau_{\kappa_\ell},$
and $U(\cdot,\cdot,\cdot)$ denotes Tricomi's confluent hypergeometric function. The quantities $\Xi^{(b_\ell)}_{z_\ell;\kappa_\ell}$, $s_{\kappa_\ell}$, and $\tau_{\kappa_\ell}$ are defined explicitly in Appendix~\ref{Appendix_C}. The detailed derivation of Eq.~\eqref{eq:QMSD_metric_scf} is provided in Appendix~\ref{Appendix_C}.

Assuming equal a priori probabilities over the set \(\{0,1\}^{L}\), the corresponding sequence error probability is given by
\begin{equation}
P_{\mathrm{seq}}^{(L)}
=
1-\frac{1}{2^L}
\sum_{\mathbf b\in\{0,1\}^{L}}
\sum_{\mathbf z\in\mathcal D(\mathbf b)}
\Lambda(\mathbf z\mid \mathbf b,\delta),
\label{eq:QMSD_seq_error}
\end{equation}
where the decision region associated with the sequence \(\mathbf b\) is defined by
\begin{equation}
\mathcal D(\mathbf b)
=
\left\{
\mathbf z:
\Lambda(\mathbf z\mid \mathbf b,\delta)
\ge
\Lambda(\mathbf z\mid \mathbf b',\delta),
\ \forall\,\mathbf b'\neq\mathbf b
\right\}.
\label{eq:QMSD_decision_region}
\end{equation}

When performance is measured in terms of bit errors over the observation
block, the accepted-only QBER of QMSD is obtained by averaging the
Hamming distance \(d_H(\cdot,\cdot)\) between the detected and transmitted sequences over the accepted conditional distribution:
\begin{equation}
\mathrm{QBER}_{\mathrm{QMSD,acc}}
=
\frac{1}{L\,2^L}
\sum_{\mathbf b\in\{0,1\}^{L}}
\sum_{\mathbf z}
d_H\!\bigl(\hat{\mathbf b}_s(\mathbf z),\mathbf b\bigr)\,
\Lambda(\mathbf z\mid \mathbf b,\delta),
\label{eq:QMSD_QBER_final}
\end{equation}
where the factor \(1/L\) accounts for the normalization by the sequence length.

\subsection{Homodyne Detection (HD)}
\label{subsec_HD}


Having analyzed QMLD and QMSD for the displaced PNR receiver, we introduce homodyne detection (HD) 
as a benchmark scheme for performance comparison.
For in-phase HD, the measured quadrature is modeled as
\begin{equation}
Y=\sqrt{I_p I_t}\,\mu\,x_A + W,
\qquad
W\sim\mathcal N(0,\sigma_H^2),
\label{eq:HD_model}
\end{equation}
where $\sigma_H^2$ denotes the effective homodyne noise variance. The corresponding binary decision is
\begin{equation}
\hat b_{\mathrm{HD}}
=
\begin{cases}
0, & Y\ge 0,\\[3pt]
1, & Y<0.
\end{cases}
\label{eq:HD_rule}
\end{equation}

The conditional HD error probability is
\begin{equation}
P_{e,\mathrm{HD}}(x_A,I_t)
=
Q\!\left(
\frac{\sqrt{I_p I_t}\,\mu\,|x_A|}{\sigma_H}
\right), \qquad \text{with}\quad Q(u)=\frac{1}{\sqrt{2\pi}}\int_u^\infty e^{-t^2/2}\,dt.
\label{eq:HD_conditional_error}
\end{equation}

The accepted-only HD QBER is obtained by averaging Eq.~\eqref{eq:HD_conditional_error} over the accepted source distribution and the fading statistics,
\begin{equation}
\mathrm{QBER}_{\mathrm{HD,acc}}
=
\frac{1}{P_{\mathrm{acc}}}
\int_{\mathbb C}\int_0^\infty
Q_m(\gamma)\,p_G(\gamma)\,
Q\!\left(
\frac{\sqrt{I_p I_t}\,\mu\,|x_A|}{\sigma_H}
\right)
p_{\mathrm E}(I_t)\,dI_t\,d^2\gamma.
\label{eq:HD_QBER_general}
\end{equation}

Introducing
\begin{equation}
V=\frac{\sqrt{I_p}\mu |x_A|}{\sigma_H},
\qquad
\bar Q_E(V)
=
\int_0^\infty Q(V\sqrt{I_t})\,p_{\mathrm E}(I_t)\,dI_t,
\qquad
\beta=\frac{V^2}{2\lambda_E},
\label{eq:HD_aux_defs}
\end{equation}
and using the closed-form expression of \(\bar Q_E(V)\) derived in Appendix~\ref{Appendix_D}, Eq.~\eqref{eq:HD_QBER_general} can be defined as follows
\begin{equation}
\mathrm{QBER}_{\mathrm{HD,acc}}
=
\frac{1}{P_{\mathrm{acc}}}
\int_{\mathbb C}
Q_m(\gamma)\,p_G(\gamma)\,
\bar Q_E\!\left(
\frac{\sqrt{I_p}\mu |x_A|}{\sigma_H}
\right)
d^2\gamma.
\label{eq:HD_QBER_reduced}
\end{equation}
With \(\gamma=x_A+i p_A\), Eq.~\eqref{eq:HD_QBER_reduced} reduces to the following semi-closed-form expression
\begin{equation}
\mathrm{QBER}_{\mathrm{HD,acc}}
=
C_{\mathrm{HD}}
\sum_{r=0}^{m}
\binom{m}{r}
\frac{\Gamma\!\left(r+\frac12\right)}{w^{r+1/2}}
\int_0^\infty
x_A^{2(m-r)}e^{-w x_A^2}
\bar Q_E\!\left(\frac{\sqrt{I_p}\mu x_A}{\sigma_H}\right)\,dx_A,
\label{eq:HD_QBER_final}
\end{equation}
where
\begin{equation}
C_{\mathrm{HD}}
=
\frac{2y^m}{m!\,\pi\,(V_T+1)\,P_{\mathrm{acc}}}.
\label{eq:CHD_def}
\end{equation}
The detailed derivation of Eq.~\eqref{eq:HD_QBER_final} is presented in Appendix~\ref{Appendix_D}.

\section{Results and Discussion}
\label{Results}
In this section, the QBER performance of the proposed VPS-QMSD scheme is evaluated and compared with two schemes: VPS-QMLD and VPS-HD. The analysis is carried out under various underwater channel conditions, including different water types (clear and coastal water), oceanic turbulence, path loss, and thermal noise at the quantum receiver.
The analytical QBER expressions derived in Section \ref{QBER} are validated by Monte Carlo simulations. For each point shown in the figures, 3000 independent transmissions are considered, with the quantum source located at a distance ``d'' from the receiver. The receiver detectors are assumed to be ideal, with unit detection efficiency throughout the performance evaluation.
The underwater turbulence-induced irradiance fluctuations are modeled using the Erlang distribution. Following the approximation reported in the literature, we consider $\theta = \lambda = 3$ and $\theta = \lambda = 10$, as these parameter values allow the Erlang distribution to closely approximate the log-normal distribution for different turbulence regimes \cite{dong2025coherent}.
The theoretical and numerical results presented below are based on a subset of the parameters given in Table~\ref{Table1}.

\begin{table}[h]
\caption{System and channel parameters.}\label{Table1}%
\centering
\small

\begin{tabular}{@{}p{0.12\textwidth}p{0.48\textwidth}p{0.27\textwidth}@{}}
\toprule
\textbf{Parameter} & \textbf{Definition} & \textbf{Value} \\
\midrule

$N$
& Thermal noise
& $0.001 \sim 1$ \cite{yuan2020free} \\

$L$
& Length of the quantum sequence
& $4, 8, 10, 12$ \\

$m$
& Number of virtually subtracted photons
& $0, 1, 2, 3$ \\

$\lambda$
& Erlang rate parameter
& $1 \sim 12$ \cite{papoulis1991probability} \\

$\theta$
& Erlang shape parameter
& $1 \sim 12$ \cite{papoulis1991probability} \\

$T$
& Transmittance of BS1
& $0.95$ \cite{hu2020continuous} \\

$\zeta$
& TMSV parameter
& $0.85$ \cite{hu2020continuous} \\

\midrule

$c$
& Extinction coefficient
& \cite{mobley1994light} \\

& Clear ocean water
& $0.151\,\mathrm{m}^{-1}$ \\

& Coastal ocean water
& $0.339\,\mathrm{m}^{-1}$ \\

\botrule
\end{tabular}
\end{table}

\newpage
\begin{center}

\begin{minipage}{0.48\textwidth}
    \centering
    \includegraphics[width=\linewidth]{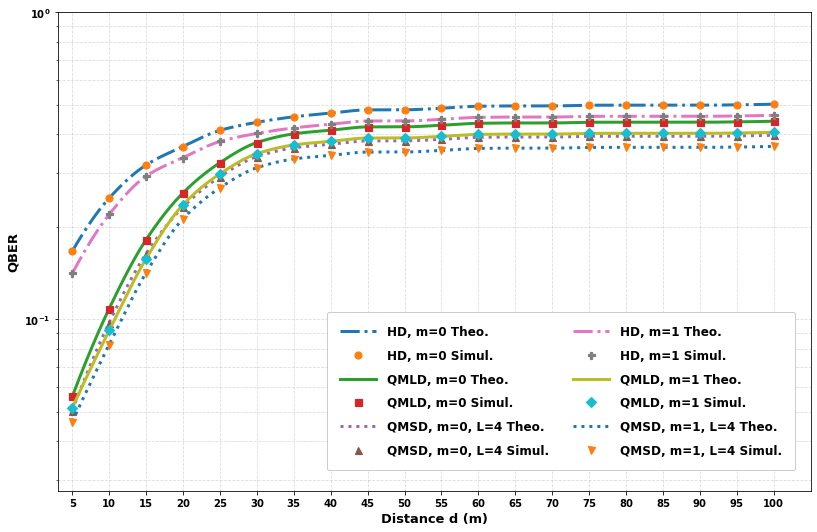}
    \textbf{(a)}
\end{minipage}
\hfill
\begin{minipage}{0.48\textwidth}
    \centering
    \includegraphics[width=\linewidth]{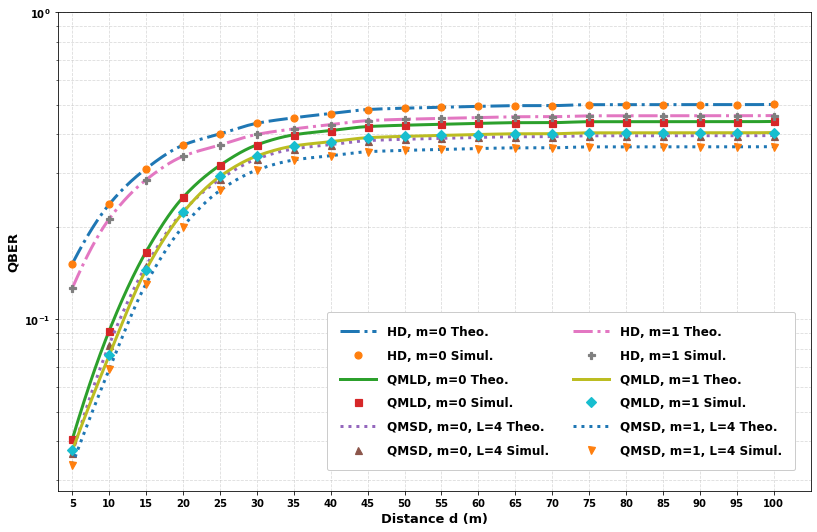}
    \textbf{(b)}
\end{minipage}

\vspace{0.3cm}

\begin{minipage}{0.48\textwidth}
    \centering
    \includegraphics[width=\linewidth]{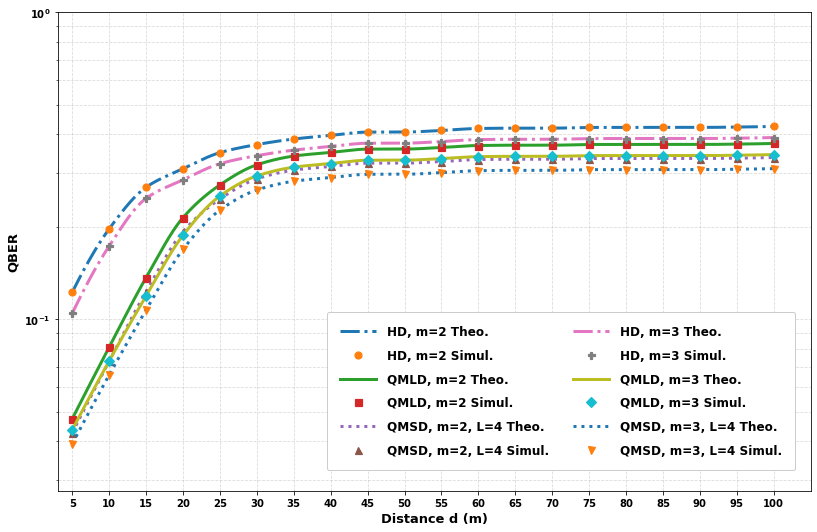}
    \textbf{(c)}
\end{minipage}
\hfill
\begin{minipage}{0.48\textwidth}
    \centering
    \includegraphics[width=\linewidth]{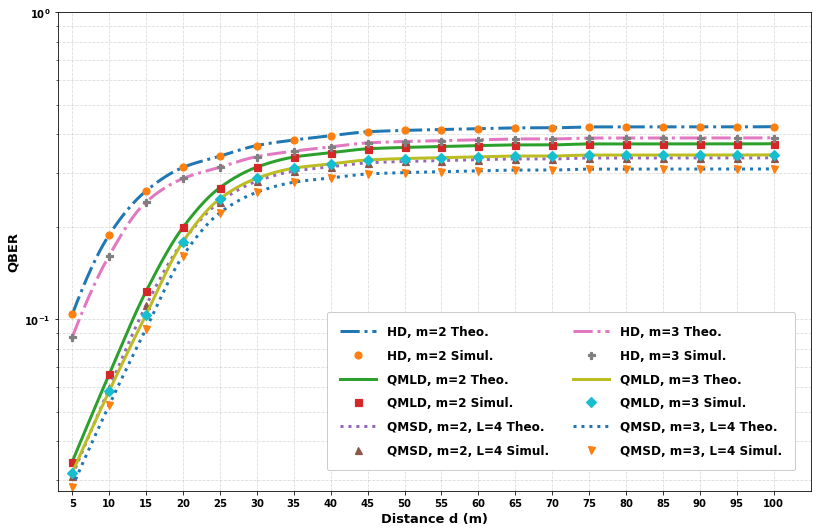}
    \textbf{(d)}
\end{minipage}

\captionof{figure}{QBER vs. distance for VPS-HD, VPS-QMLD, and VPS-QMSD over clear ocean water, under different marine turbulence conditions and different numbers of virtually subtracted photons $m$: (a) $\lambda=\theta=3$, $m=0$ and $m=1$; (b) $\lambda=\theta=10$, $m=0$ and $m=1$; (c) $\lambda=\theta=3$, $m=2$ and $m=3$; and (d) $\lambda=\theta=10$, $m=2$ and $m=3$.}
\label{Figure-1-}

\end{center}

Fig.~\ref{Figure-1-} presents a comparison of the QBER performance between the proposed VPS-QMSD scheme, VPS-QMLD, and VPS-HD as a function of the transmission distance d over a clear underwater channel with low thermal noise ($N=0.001$). The evaluation considers a quantum sequence length L=4 under various marine turbulence conditions ($\theta = \lambda = 3$ and $\theta = \lambda = 10$) and photon-subtraction orders $m$. The excellent agreement between the analytical curves and Monte Carlo simulation points across all considered scenarios validates the theoretical framework and confirms the accuracy of the derived expressions. As expected, the QBER increases with the transmission distance for all configurations, reflecting the progressive degradation of the optical field during propagation. Notably, the proposed VPS-QMSD scheme consistently achieves the lowest QBER, followed by VPS-QMLD, whereas VPS-HD exhibits the poorest performance. This behavior demonstrates the clear advantage of sequence-level quantum detection, which effectively exploits the channel's spatial-temporal correlations without requiring instantaneous CSI.

In Fig.~\ref{Figure-1-}(a), which corresponds to severe turbulence ($\theta = \lambda = 3$) with $m=0$ and $m=1$, implementing a single-photon virtual subtraction ($m=1$) yields a lower QBER than the non-subtracted case ($m=0$) across all receiver structures. This confirms that the non-Gaussian state engineering enabled by VPS successfully mitigates channel-induced degradation, thereby enhancing detection reliability. In Fig.~\ref{Figure-1-}(b), obtained under weaker turbulence conditions ($\theta = \lambda = 10$), the same performance hierarchy is preserved, but with significantly lower absolute QBER values. This improvement stems from the reduced irradiance fluctuations associated with larger Erlang parameters, which naturally weakens the impact of fading.

A similar trend is observed for higher photon-subtraction orders. In Fig.~\ref{Figure-1-}(c) \mbox{($\theta = \lambda = 3$, $m=2$ and $m=3$),} increasing the subtraction order further reduces the QBER, with $m=3$ outperforming $m=2$, which in turn remains superior to the $m=1$ case. Finally, Fig.~\ref{Figure-1-}(d) illustrates that the joint benefit of a less turbulent medium ($\theta = \lambda = 10$) and higher VPS orders ($m=2,3$) provides the most robust operating regime. This confirms that the performance gains unlocked by combining VPS and QMSD remain highly significant across all channel conditions.

\begin{center}
\includegraphics[width=1.0\textwidth]{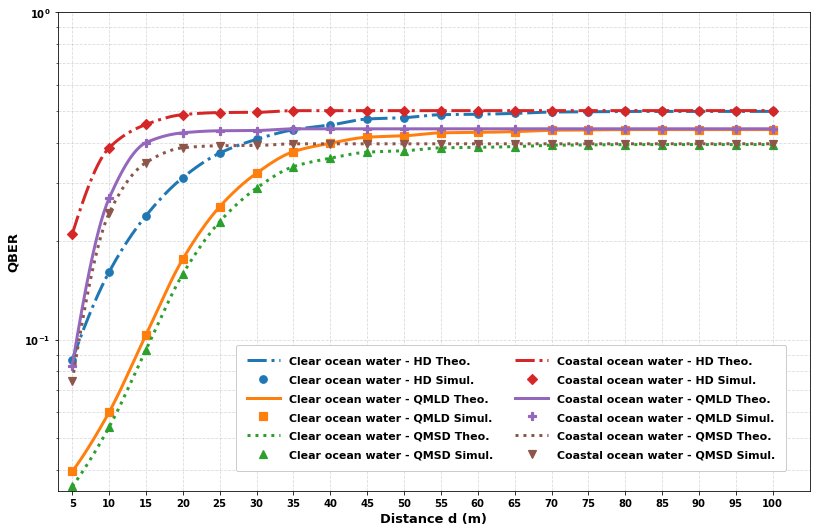}
\captionof{figure}{QBER vs. distance for VPS-HD, VPS-QMLD, and VPS-QMSD in clear ocean water and coastal ocean water, with $N=0.001$, $\lambda=\theta=10$, $m=3$, $L=4$.
\label{Figure-2-}}
\end{center}

In the following analysis, we assess the QBER performance of the proposed VPS-QMSD scheme in comparison with VPS-QMLD and VPS-HD for clear ocean water and coastal ocean water, as depicted in Fig.~\ref{Figure-2-}. For both water types, VPS-QMSD achieves the lowest QBER, followed by VPS-QMLD, whereas VPS-HD gives the highest error rate, which confirms the effectiveness of multiple-symbol detection under underwater propagation conditions. A distinct difference appears between the two water types. For a given transmission distance, all three schemes provide lower QBER in clear ocean water than in coastal ocean water. This behavior can be attributed to the less severe attenuation, scattering, and turbulence conditions in clear ocean water than in coastal ocean water, which results in better signal preservation and improved detection reliability. In contrast, coastal ocean water is associated with a steeper increase in QBER with transmission distance than clear ocean water, owing to stronger attenuation and scattering effects.

\begin{center}
\includegraphics[width=1.0\textwidth]{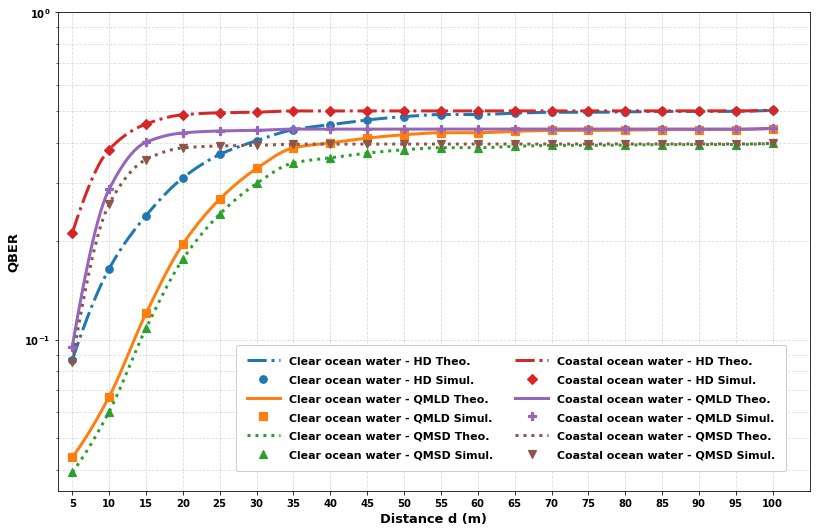}
\captionof{figure}{QBER vs. distance for VPS-HD, VPS-QMLD, and VPS-QMSD in clear ocean water and coastal ocean water, with $N=1$, $\lambda=\theta=10$, $m=3$, and $L=4$.
\label{Figure-3-}}
\end{center}

Fig.~\ref{Figure-3-} shows the impact of a higher thermal noise level on the QBER performance of VPS-HD, VPS-QMLD, and VPS-QMSD in clear ocean water and coastal ocean water, with $N=1$, $m=3$, and $\lambda=\theta=10$, in comparison with Fig.~\ref{Figure-2-}. The increase in thermal noise leads to higher QBER values for all three schemes and for both water types over the entire transmission distance range. This degradation is caused by the stronger receiver thermal noise, which reduces the effective SNR and makes the received quantum signal more vulnerable to decision errors. Despite this performance loss, the same ordering among the three schemes is preserved, with VPS-QMSD providing the lowest QBER, followed by VPS-QMLD and VPS-HD. The difference between clear ocean water and coastal ocean water also remains visible, as clear ocean water still provides better performance than coastal ocean water.

\begin{center}
\includegraphics[width=1.0\textwidth]{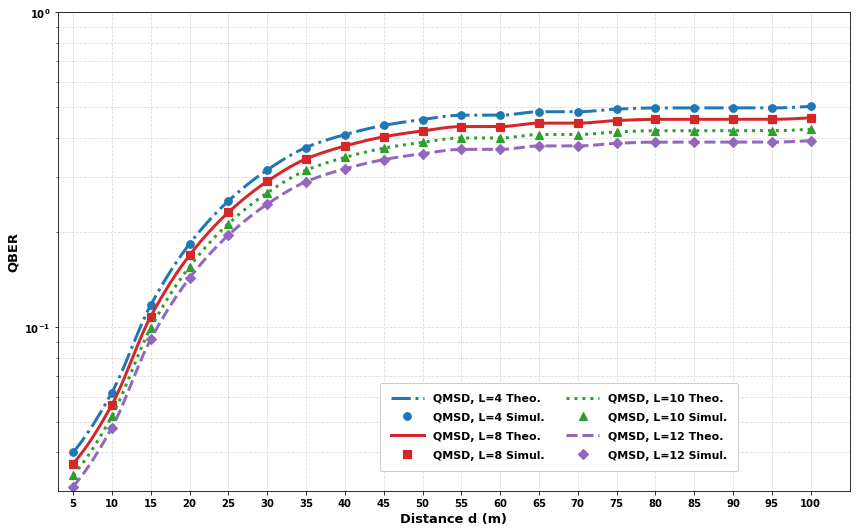}
\captionof{figure}{QBER vs. distance VPS-QMSD scheme with different quantum sequence lengths $L$, in clear ocean water, under identical channel gain parameters $\lambda=\theta=10$, $m=3$, and thermal noise \mbox{$N=0.001$.}
\label{Figure-4-}}
\end{center}

Fig.~\ref{Figure-4-} illustrates the impact of the quantum sequence length $L$ on the QBER performance of the VPS-QMSD scheme as the transmission distance increases. For all considered distances, the QBER decreases as $L$ increases, with the best performance achieved for $L=12$, followed by $L=10$, $L=8$, and $L=4$. This behavior is explained by the fact that QMSD performs joint detection over a block of consecutive received symbols rather than processing each symbol independently. By increasing the sequence length, the receiver exploits a larger set of observations and makes better use of the statistical correlation of the underwater turbulent channel over the observation window. As a result, the decision process becomes less sensitive to instantaneous channel fluctuations, since the effect of fading is effectively averaged over a longer sequence. This leads to a more reliable detection metric and, consequently, to a lower QBER. Increasing the quantum sequence length therefore improves the robustness of the VPS-QMSD scheme and provides a more effective reduction in QBER under the considered underwater channel conditions.

The QBER values reported in all figures are interpreted within the more constrained operating scenario of the considered system model, which jointly accounts for the main transmitter, channel, and receiver impairments. Therefore, rather than directly comparing QBER values with those obtained under simplified models that consider VPS or QMSD separately, the analysis highlights the contribution of each receiver scheme within the same modeling framework.

\section{Conclusion}
\label{Conclusion}
This work presented a CSI-free underwater CV-QKD system based on virtual photon subtraction at Alice’s side and three receiver schemes at Bob: VPS-HD, VPS-QMLD, and VPS-QMSD. The underwater channel was comprehensively modeled to account for absorption, scattering, Erlang-modeled turbulence, path loss, and receiver thermal noise. Since VPS retains only the accepted states, the system performance was evaluated using the accepted-only QBER.

Analytical and semi-closed-form QBER expressions were derived for the considered schemes and validated through Monte Carlo simulations. The excellent agreement between theoretical predictions and simulation results confirms the accuracy of the developed framework. The results showed that the QBER increases with transmission distance, thermal noise, and more unfavorable propagation conditions. Clear ocean water provided better performance than coastal ocean water due to lower attenuation and scattering.

The analysis also highlights the complementary roles of transmitter-side VPS and receiver-side sequence detection. VPS reshapes the accepted source distribution and improves the reliability of the retained states, whereas QMSD enhances the decision process by exploiting block-level information from the received photon-count observations. The synergy of these two techniques provides a stronger performance gain than employing VPS with a symbol-by-symbol detection scheme or homodyne detection.

Among the three considered schemes, VPS-QMSD consistently achieved the lowest QBER in all investigated cases, followed by VPS-QMLD and VPS-HD. This improvement is mainly due to the ability of QMSD to jointly process a block of observations and exploit the statistical correlation of the underwater turbulent channel without requiring instantaneous CSI at the receiver. The results also showed that increasing the VPS order and the sequence length improves system robustness and further reduces the QBER.

These findings confirm that combining VPS-based source post-selection with CSI-free multiple-symbol detection is an effective approach for reducing the QBER of underwater CV-QKD systems under turbulence, path loss, and receiver thermal noise.

\section{Acknowledgments}
The authors sincerely acknowledge the financial support provided by the Brittany Region for this research.
\section*{Declarations}

\subsection*{Conflict of interest}

The authors declare that they have no conflict of interest.

\subsection*{Data availability}

No external datasets were used in this study. The data generated through
numerical simulations based on the models and parameters described in the
manuscript are available from the corresponding author upon reasonable request.
\appendix
\renewcommand{\thesection}{Appendix \Alph{section}}
\section{Semi-Closed-Form Representations of $\Pr(z\mid b,I_t,\delta)$}
\label{Appendix_A}
\setcounter{equation}{0}
\renewcommand{\theequation}{A.\arabic{equation}}
Starting from Eq.~\eqref{eq:PrzbItd_expanded}, the phase-space variable is written in polar form as
\begin{equation}
\gamma = r e^{i\phi},
\qquad
d^2\gamma = r\,dr\,d\phi,
\qquad
\phi \in (0,2\pi].
\label{eq:A_polar}
\end{equation}
Since the binary mapping is defined by the sign of $x_A=\Re\{\gamma\}$, the integration domain reduces to the angular region $\Phi_b$, with
\begin{equation}
\Phi_0=\left[-\frac{\pi}{2},\frac{\pi}{2}\right),
\qquad
\Phi_1=\left[\frac{\pi}{2},\frac{3\pi}{2}\right).
\label{eq:A_phi_domains}
\end{equation}
From the polar representation in Eq.~\eqref{eq:A_polar}, Eq.~\eqref{eq:Sdelta} is represented as
\begin{equation}
S_\delta(r,\phi,I_t)
=
I_p I_t\mu^2 r^2+|\delta|^2
-2|\delta|\,\mu\sqrt{I_pI_t}\,r\cos(\phi-\phi_\delta).
\label{eq:A_SdeltaPolar}
\end{equation}

Using Eqs.~\eqref{eq:pG} and \eqref{eq:Qm}, and noting that $|\gamma|^2=r^2$, their product reduces to
\begin{equation}
Q_m(r)\,p_G(r)
=
\frac{y^m}{m!\,\pi(V_T+1)}\,r^{2m}e^{-wr^2},
\qquad
w=y+\frac{1}{V_T+1}.
\label{eq:A_QpG}
\end{equation}

By substituting Eqs.~\eqref{eq:A_polar} - \eqref{eq:A_QpG} and Eq.~\eqref{eq:PNR} into Eq.~\eqref{eq:PrzbItd_expanded}, and using Eq.~\eqref{eq:Pacc_bit_symmetry} in the final line, one obtains Eq.~\eqref{eq:PrzbItd_angular_exact}
{\small
\begin{align}
\Pr(z\mid b,I_t,\delta)
&=
\frac{1}{P_{\mathrm{acc}}^{(b)}}
\int_0^\infty\int_{\Phi_b}
\Pr(z\mid r,\phi,I_t,\delta)\,
Q_m(r)\,p_G(r)\,r\,d\phi\,dr
\notag\\
&=
\frac{1}{P_{\mathrm{acc}}^{(b)}}
\int_0^\infty\int_{\Phi_b}
\frac{N^z}{(N+1)^{z+1}}
\exp\!\left(-\frac{S_\delta}{N+1}\right)
L_z\!\left(-\frac{S_\delta}{N(N+1)}\right)\notag\\
&\times \frac{y^m}{m!\,\pi(V_T+1)}\,r^{2m}e^{-wr^2}\,
r\,d\phi\,dr
\notag\\
&=
\frac{2y^m N^z}{m!\,\pi(V_T+1)\,P_{\mathrm{acc}}\,(N+1)^{z+1}}
\int_0^\infty r^{2m+1}e^{-wr^2}\notag\\
& \times \left[
\int_{\Phi_b}
e^{-S_\delta/(N+1)}
L_z\!\left(-\frac{S_\delta}{N(N+1)}\right)\,d\phi
\right]dr.
\label{eq:A_exact_final}
\end{align}}

Combining Eq.~\eqref{eq:LaguerreExpand} with Eq.~\eqref{eq:A_exact_final} gives
\begin{equation}
\Pr(z\mid b,I_t,\delta)
=
\frac{N^z}{(N+1)^{z+1}}
\sum_{k=0}^{z}
\binom{z}{k}
\frac{1}{k!\,[N(N+1)]^k}
V_{k,b}(I_t,\delta),
\label{eq:A_k_sum}
\end{equation}
where
\begin{equation}
V_{k,b}(I_t,\delta)
=
\frac{2y^m}{m!\,\pi(V_T+1)\,P_{\mathrm{acc}}}
\int_0^\infty
r^{2m+1}e^{-wr^2}
\left[
\int_{\Phi_b}
e^{-S_\delta/(N+1)}S_\delta^k\,d\phi
\right]dr.
\label{eq:A_Vkb}
\end{equation}

When the displacement phase is aligned with the in-phase quadrature, i.e., $\phi_\delta=0$, Eq.~\eqref{eq:A_SdeltaPolar} becomes
\begin{equation}
S_\delta(r,\phi,I_t)=a(r,I_t)-c(r,I_t)\cos\phi,
\label{eq:A_Sacos}
\end{equation}
with
\begin{equation}
a(r,I_t)=I_p I_t\mu^2r^2+|\delta|^2,
\qquad
c(r,I_t)=2|\delta|\,\mu\sqrt{I_pI_t}\,r,
\qquad
\eta(r,I_t)=\frac{c(r,I_t)}{N+1}.
\label{eq:A_params}
\end{equation}
Hence,
\begin{equation}
e^{-S_\delta/(N+1)}=e^{-a/(N+1)}e^{\eta\cos\phi},
\label{eq:A_exp_split}
\end{equation}
and
\begin{equation}
S_\delta^k=(a-c\cos\phi)^k
=
\sum_{n=0}^{k}\binom{k}{n}a^{k-n}(-c)^n\cos^n\phi.
\label{eq:A_poly_expand}
\end{equation}

Substituting Eqs.~\eqref{eq:A_exp_split} and \eqref{eq:A_poly_expand} into Eq.~\eqref{eq:A_Vkb} yields
\begin{equation}
V_{k,b}(I_t,\delta)
=
C_0
\int_0^\infty
r^{2m+1}e^{-wr^2}e^{-a/(N+1)}
\sum_{n=0}^{k}\binom{k}{n}a^{k-n}(-c)^n
\mathcal{B}_n^{(b)}(\eta)\,dr,
\label{eq:A_Vkb_Bn}
\end{equation}
where $C_0$ is defined using Eq.~\eqref{eq:Pacc_bit_symmetry} as
\begin{equation}
C_0=
\frac{2y^m}{m!\,\pi(V_T+1)\,P_{\mathrm{acc}}},
\label{eq:A_C0}
\end{equation}
and the bit-dependent angular moments are given by
\begin{equation}
\mathcal{B}_n^{(b)}(\eta)
=
\int_{\Phi_b}e^{\eta\cos\phi}\cos^n\phi\,d\phi.
\label{eq:A_Bnb}
\end{equation}

Introducing the half-range moments
\begin{equation}
\mathcal{B}_n(\eta)
=
\int_{-\pi/2}^{\pi/2}e^{\eta\cos\phi}\cos^n\phi\,d\phi,
\label{eq:A_Bn}
\end{equation}
Eq.~\eqref{eq:A_Bnb} satisfies
\begin{equation}
\mathcal{B}_n^{(0)}(\eta)=\mathcal{B}_n(\eta),
\qquad
\mathcal{B}_n^{(1)}(\eta)=(-1)^n\mathcal{B}_n(-\eta).
\label{eq:A_Bn_rel}
\end{equation}

Expanding \(e^{\eta\cos\phi}\) into a power series, Eq.~\eqref{eq:A_Bn} can be expressed as
\begin{equation}
\mathcal{B}_n(\eta)
=
\sum_{j=0}^{\infty}\frac{\eta^j}{j!}M_{n+j},
\label{eq:A_Bn_series}
\end{equation}
where
\begin{equation}
M_q
=
\int_{-\pi/2}^{\pi/2}\cos^q\phi\,d\phi
=
\sqrt{\pi}\,
\frac{\Gamma\!\left(\frac{q+1}{2}\right)}
{\Gamma\!\left(\frac{q+2}{2}\right)},
\label{eq:A_Mq}
\end{equation}
and $\Gamma(\cdot)$ denotes the Gamma function.

Therefore, Eq.~\eqref{eq:A_Bnb} reduces to
\begin{equation}
\mathcal{B}_n^{(b)}(\eta)
=
\sum_{j=0}^{\infty}
\frac{\eta^j}{j!}\,
S_b(n,j)\,
M_{n+j},
\label{eq:A_Bnb_series}
\end{equation}
with
\begin{equation}
S_b(n,j)=
\begin{cases}
1, & b=0,\\[3pt]
(-1)^{n+j}, & b=1.
\end{cases}
\label{eq:A_Sb}
\end{equation}

Substituting Eq.~\eqref{eq:A_Bnb_series} into Eq.~\eqref{eq:A_Vkb_Bn}, and then the result into Eq.~\eqref{eq:A_k_sum}, yields a semi-closed-form of \(\Pr(z\mid b,I_t,\delta)\).

\section[\appendixname~\thesection]{Semi-Closed-Form for the CSI-free likelihood}
\label{Appendix_B}
\setcounter{equation}{0}
\renewcommand{\theequation}{B.\arabic{equation}}

Starting from Eq. \eqref{eq:Przbd}, the semi-closed-form expression is obtained by inserting the semi-closed-form representation of \(\Pr(z\mid b,I_t,\delta)\) derived in Appendix~A. This gives

\begin{align}
\Pr(z\mid b,\delta)
&=
\int_0^\infty
\frac{N^z}{(N+1)^{z+1}}
\sum_{k=0}^{z}\binom{z}{k}\frac{1}{k!\,[N(N+1)]^k}
\,C_0
\int_0^\infty
r^{2m+1}e^{-wr^2}e^{-a/(N+1)}
\notag\\
&\qquad\qquad\qquad\times
\sum_{n=0}^{k}\binom{k}{n}a^{k-n}(-c)^n
\left[
\sum_{j=0}^{\infty}\frac{\eta^j}{j!}S_b(n,j)M_{n+j}
\right]
dr\;
p_{\mathrm E}(I_t)\,dI_t.
\label{eq:B1}
\end{align}

Using
\begin{equation}
e^{-a/(N+1)}
=
\exp\!\left(-\frac{|\delta|^2}{N+1}\right)
\exp\!\left(-\frac{I_p I_t\mu^2r^2}{N+1}\right),
\label{eq:B2}
\end{equation}
together with
\begin{equation}
a^{k-n}
=
\sum_{i=0}^{k-n}
\binom{k-n}{i}
(|\delta|^2)^{k-n-i}(I_p\mu^2r^2)^i I_t^i,
\label{eq:B3}
\end{equation}
and
\begin{equation}
c^n=(2|\delta|\mu\sqrt{I_p})^n r^n I_t^{n/2},
\qquad
\eta^j=\left(\frac{2|\delta|\mu\sqrt{I_p}}{N+1}\right)^j r^j I_t^{j/2},
\label{eq:B4}
\end{equation}
the dependence on \(I_t\) can be collected explicitly. Introducing
\begin{equation}
\Omega=2|\delta|\mu\sqrt{I_p},
\qquad
\rho=\frac{\Omega}{N+1},
\qquad
\chi=\frac{I_p\mu^2}{N+1},
\label{eq:B5}
\end{equation}
and substituting Eqs. \eqref{eq:B2}--\eqref{eq:B5} into Eq. \eqref{eq:B1}, while using Eq.~\eqref{eq:Pacc_bit_symmetry}, gives
{\small
\begin{align}
\Pr(z\mid b,\delta)
&=
\frac{N^z}{(N+1)^{z+1}}
\cdot
\frac{2y^m}{m!\,\pi(V_T+1)\,P_{\mathrm{acc}}}
\cdot
\frac{\lambda_E^\theta}{(\theta-1)!}
\cdot
\exp\!\left(-\frac{|\delta|^2}{N+1}\right)
\notag\\
&\quad\times
\sum_{k=0}^{z}\binom{z}{k}\frac{1}{k!\,[N(N+1)]^k}
\sum_{n=0}^{k}\binom{k}{n}(-1)^n\Omega^n
\sum_{i=0}^{k-n}\binom{k-n}{i}
(|\delta|^2)^{k-n-i}(I_p\mu^2)^i
\notag\\
&\quad\times
\sum_{j=0}^{\infty}\frac{\rho^j}{j!}S_b(n,j)M_{n+j}
\int_0^\infty\int_0^\infty
r^{2m+1+2i+n+j}
e^{-wr^2}
I_t^{\theta-1+i+\frac{n+j}{2}}
e^{-(\lambda_E+\chi r^2)I_t}
\,dI_t\,dr.
\label{eq:B6}
\end{align}}

The inner integral with respect to \(I_t\) is of a Gamma type :
\begin{equation}
\int_0^\infty
I_t^{\theta-1+i+\frac{n+j}{2}}
e^{-(\lambda_E+\chi r^2)I_t}\,dI_t
=
\frac{
\Gamma\!\left(\theta+i+\frac{n+j}{2}\right)
}{
(\lambda_E+\chi r^2)^{\theta+i+\frac{n+j}{2}}
}.
\label{eq:B7}
\end{equation}
Substituting Eq. \eqref{eq:B7} into Eq. \eqref{eq:B6}, and defining
\begin{equation}
\mathcal{K}_z
=
\frac{N^z}{(N+1)^{z+1}}
\cdot
\frac{2y^m}{m!\,\pi(V_T+1)\,P_{\mathrm{acc}}}
\cdot
\frac{\lambda_E^\theta}{(\theta-1)!}
\cdot
\exp\!\left(-\frac{|\delta|^2}{N+1}\right),
\label{eq:B8}
\end{equation}
one obtains Eq. \eqref{eq:Przbd_semiclosed}
\begin{align}
\Pr(z\mid b,\delta)
&=
\mathcal{K}_z
\sum_{k=0}^{z}\binom{z}{k}\frac{1}{k!\,[N(N+1)]^k}
\sum_{n=0}^{k}\binom{k}{n}(-1)^n\Omega^n
\sum_{i=0}^{k-n}\binom{k-n}{i}
(|\delta|^2)^{k-n-i}(I_p\mu^2)^i
\notag\\
&\quad\times
\sum_{j=0}^{\infty}\frac{\rho^j}{j!}S_b(n,j)M_{n+j}
\mathcal{I}_{k,n,i,j},
\label{eq:B9}
\end{align}
where
\begin{equation}
\mathcal{I}_{k,n,i,j}
=
\int_0^\infty
r^{2m+1+2i+n+j}
e^{-wr^2}
\frac{
\Gamma\!\left(\theta+i+\frac{n+j}{2}\right)
}{
(\lambda_E+\chi r^2)^{\theta+i+\frac{n+j}{2}}
}
\,dr.
\label{eq:B10}
\end{equation}

The remaining radial integral admits a closed-form representation in terms of Tricomi's confluent hypergeometric function. Letting \(u=r^2\), one obtains
\begin{equation}
\mathcal{I}_{k,n,i,j}
=
\frac{1}{2}\Gamma(\nu)
\int_0^\infty
u^{\tau-1}e^{-wu}(\lambda_E+\chi u)^{-\nu}\,du,
\label{eq:B11}
\end{equation}
with
\begin{equation}
\tau=m+i+\frac{n+j}{2}+1,
\qquad
\nu=\theta+i+\frac{n+j}{2}.
\label{eq:B12}
\end{equation}
Using the integral representation of Tricomi's function gives
\begin{equation}
\mathcal{I}_{k,n,i,j}
=
\frac{1}{2}
\Gamma(\nu)\Gamma(\tau)
\chi^{-\tau}\lambda_E^{\tau-\nu}
U\!\left(
\tau,\,\tau+1-\nu,\,\frac{\lambda_E w}{\chi}
\right).
\label{eq:B13}
\end{equation}

\section[\appendixname~\thesection]{Semi-Closed-Form for the CSI-free QMSD Metric}
\label{Appendix_C}
\setcounter{equation}{0}
\renewcommand{\theequation}{C.\arabic{equation}}

Using the semi-closed-form representation of the conditional symbol likelihood established in Eq.\eqref{eq:A_k_sum}, Eq.~\eqref{eq:QMSD_metric} is evaluated by rewriting each term \(\Pr(z\mid b,I_t,\delta)\) in a separable form with respect to \(I_t\)
\begin{equation}
\Pr(z\mid b,I_t,\delta)
=
\sum_{k=0}^{z}
\sum_{n=0}^{k}
\sum_{i=0}^{k-n}
\sum_{j=0}^{\infty}
\Xi^{(b)}_{z;k,n,i,j}\,
I_t^{\,s}\,
(w+\chi I_t)^{-\tau},
\label{eq:C2}
\end{equation}
where
\begin{equation}
\left\{
\begin{aligned}
s&=i+\frac{n+j}{2},\hspace{1cm}
\tau=m+i+\frac{n+j}{2}+1, \hspace{1cm}
\chi=\frac{I_p\mu^2}{N+1},\\
\Omega&=2|\delta|\mu\sqrt{I_p},\hspace{1cm}
\rho=\frac{\Omega}{N+1}.
\end{aligned}
\right.
\label{eq:C3}
\end{equation}

The coefficient \(\Xi^{(b)}_{z;k,n,i,j}\) collects all terms independent of \(I_t\). Using Eq.~\eqref{eq:Pacc_bit_symmetry}, it is given by
\begin{align}
\Xi^{(b)}_{z;k,n,i,j}
&=
\frac{2y^m}{m!\,\pi\,(V_T+1)\,P_{\mathrm{acc}}}
\frac{N^z}{(N+1)^{z+1}}
\binom{z}{k}
\frac{1}{k!\,[N(N+1)]^k}
\binom{k}{n}
(-1)^n\Omega^n\notag \\
&\times\binom{k-n}{i}
|\delta|^{2(k-n-i)}
(I_p\mu^2)^i\times
\frac{\rho^j}{j!}\,
S_b(n,j)\,
M_{n+j}\,
\frac{\Gamma(\tau)}{2}\,
\exp\!\left(-\frac{|\delta|^2}{N+1}\right).
\label{eq:C5}
\end{align}

Eq.~\eqref{eq:C2} follows from the radial integral
\begin{equation}
\int_0^\infty
r^{2m+1+2i+n+j}
e^{-(w+\chi I_t)r^2}\,dr
=
\frac{1}{2}\Gamma(\tau)(w+\chi I_t)^{-\tau}.
\label{eq:C6}
\end{equation}

Substituting Eq.~\eqref{eq:C2} into Eq.~\eqref{eq:QMSD_metric} yields
\begin{equation}
\Lambda(\mathbf z\mid \mathbf b,\delta)
=
\int_0^\infty
\prod_{\ell=1}^{L}
\left[
\sum_{\kappa_\ell}
\Xi^{(b_\ell)}_{z_\ell;\kappa_\ell}\,
I_t^{s_{\kappa_\ell}}
(w+\chi I_t)^{-\tau_{\kappa_\ell}}
\right]
p_{\mathrm E}(I_t)\,dI_t,
\label{eq:C7}
\end{equation}
with \(\kappa_\ell=(k_\ell,n_\ell,i_\ell,j_\ell)\).

Expanding the product term by term, Eq.~\eqref{eq:C7} becomes
\begin{equation}
\Lambda(\mathbf z\mid \mathbf b,\delta)
=
\sum_{\kappa_1}\cdots\sum_{\kappa_L}
\left[
\prod_{\ell=1}^{L}
\Xi^{(b_\ell)}_{z_\ell;\kappa_\ell}
\right]
\int_0^\infty
I_t^{\Sigma_s}
(w+\chi I_t)^{-\Sigma_\tau}
p_{\mathrm E}(I_t)\,dI_t,
\label{eq:C9}
\end{equation}
with
\begin{equation}
\Sigma_s=\sum_{\ell=1}^{L}s_{\kappa_\ell},
\qquad
\Sigma_\tau=\sum_{\ell=1}^{L}\tau_{\kappa_\ell}.
\label{eq:C10}
\end{equation}

Using the Erlang density in Eq.~\eqref{eq:erlang_pdf}, the remaining integral in Eq.~\eqref{eq:C9} becomes
\begin{equation}
\mathcal J(\Sigma_s,\Sigma_\tau)
=
\frac{\lambda_E^\theta}{\Gamma(\theta)}
\int_0^\infty
I_t^{\theta+\Sigma_s-1}
e^{-\lambda_E I_t}
(w+\chi I_t)^{-\Sigma_\tau}
\,dI_t.
\label{eq:C12}
\end{equation}

Factoring \(w\) in Eq.~\eqref{eq:C12} gives
\begin{equation}
(w+\chi I_t)^{-\Sigma_\tau}
=
w^{-\Sigma_\tau}
\left(1+\frac{\chi}{w}I_t\right)^{-\Sigma_\tau},
\label{eq:C13}
\end{equation}
and applying the change of variable $x=\frac{\chi}{w}I_t,
\,\,
I_t=\frac{w}{\chi}x,
\,\,
dI_t=\frac{w}{\chi}\,dx,$ one obtains
\begin{equation}
\mathcal J(\Sigma_s,\Sigma_\tau)
=
\frac{\lambda_E^\theta}{\Gamma(\theta)}
\chi^{-(\theta+\Sigma_s)}
w^{\theta+\Sigma_s-\Sigma_\tau}
\int_0^\infty
x^{\theta+\Sigma_s-1}
e^{-(\lambda_E w/\chi)x}
(1+x)^{-\Sigma_\tau}
\,dx.
\label{eq:C15}
\end{equation}

Using the integral representation of Tricomi's confluent hypergeometric function
\begin{equation}
U(a,b,\psi)
=
\frac{1}{\Gamma(a)}
\int_0^\infty
t^{a-1}e^{-\psi t}(1+t)^{b-a-1}\,dt,
\label{eq:C16}
\end{equation}
with 
$a=\theta+\Sigma_s,
\,\,
b=\theta+\Sigma_s+1-\Sigma_\tau,
\,\,
\psi=\frac{\lambda_E w}{\chi},$ Eq.~\eqref{eq:C15} reduces to
\begin{equation}
\mathcal J(\Sigma_s,\Sigma_\tau)
=
\frac{\lambda_E^\theta}{\Gamma(\theta)}
\chi^{-(\theta+\Sigma_s)}
w^{\theta+\Sigma_s-\Sigma_\tau}
\Gamma(\theta+\Sigma_s)
U\!\left(
\theta+\Sigma_s,\,
\theta+\Sigma_s+1-\Sigma_\tau,\,
\frac{\lambda_E w}{\chi}
\right).
\label{eq:C18}
\end{equation}

Finally, substituting Eq.~\eqref{eq:C18} into Eq.~\eqref{eq:C9}, one obtains
\begin{align}
\Lambda(\mathbf z\mid \mathbf b,\delta)
&=
\frac{\lambda_E^\theta}{\Gamma(\theta)}
\sum_{\kappa_1}\cdots\sum_{\kappa_L}
\left[
\prod_{\ell=1}^{L}
\Xi^{(b_\ell)}_{z_\ell;\kappa_\ell}
\right]
\chi^{-(\theta+\Sigma_s)}
w^{\theta+\Sigma_s-\Sigma_\tau}
\Gamma(\theta+\Sigma_s)
\nonumber\\
&\quad\times
U\!\left(
\theta+\Sigma_s,\,
\theta+\Sigma_s+1-\Sigma_\tau,\,
\frac{\lambda_E w}{\chi}
\right),
\label{eq:C19}
\end{align}
which is Eq.~\eqref{eq:QMSD_metric_scf}.

\section[\appendixname~\thesection]{Semi-Closed-Form for the HD QBER Under Erlang Fading}
\label{Appendix_D}
\setcounter{equation}{0}
\renewcommand{\theequation}{D.\arabic{equation}}

From Eq.~\eqref{eq:HD_aux_defs}, we have $u=V\sqrt{I_t}\ge 0$. Hence, Craig's representation of the Gaussian $Q$-function can be applied as
\begin{equation}
Q(u)
=
\frac{1}{\pi}
\int_{0}^{\pi/2}
\exp\!\left(-\frac{u^2}{2\sin^2\phi}\right)d\phi,
\qquad u\ge 0 .
\label{eq:app_HD_Craig}
\end{equation}
Substituting Eq.~\eqref{eq:app_HD_Craig} into the expression of
$\bar Q_E(V)$ yields
\begin{equation}
\bar Q_E(V)
=
\frac{1}{\pi}
\int_{0}^{\pi/2}
\left[
\int_{0}^{\infty}
\exp\!\left(
-\frac{V^2 I_t}{2\sin^2\phi}
\right)
p_{\mathrm E}(I_t)\,dI_t
\right]d\phi .
\label{eq:app_HD_Qbar_Craig}
\end{equation}
Using the Laplace transform of the Erlang-distributed irradiance $I_t$,
with \mbox{$s=V^2/(2\sin^2\phi)$,} and $\beta=V^2/(2\lambda_E)$, Eq.~\eqref{eq:app_HD_Qbar_Craig} becomes
\begin{align}
\bar Q_E(V)
&=
\frac{1}{\pi}
\int_{0}^{\pi/2}
\left(
\frac{\lambda_E}
{\lambda_E+\frac{V^2}{2\sin^2\phi}}
\right)^{\theta}
d\phi
\nonumber\\
&=
\frac{1}{\pi}
\int_{0}^{\pi/2}
\left(
1+\frac{\beta}{\sin^2\phi}
\right)^{-\theta}
d\phi,
\label{eq:app_HD_Qbar_beta}
\end{align}
with the change of variable $t=\sin^2\phi$, one has
$d\phi=dt/(2\sqrt{t(1-t)})$. Hence,
\begin{align}
\bar Q_E(V)
&=
\frac{1}{2\pi}
\int_{0}^{1}
t^{\theta-\frac{1}{2}}
(1-t)^{-\frac{1}{2}}
(t+\beta)^{-\theta}
dt
\nonumber\\
&=
\frac{\beta^{-\theta}}{2\pi}
\int_{0}^{1}
t^{\theta-\frac{1}{2}}
(1-t)^{-\frac{1}{2}}
\left(1+\frac{t}{\beta}\right)^{-\theta}
dt.
\label{eq:app_HD_Qbar_factored}
\end{align}
Euler's integral representation of the Gauss hypergeometric function
applied to Eq.~\eqref{eq:app_HD_Qbar_factored} gives
\begin{equation}
\bar Q_E(V)
=
\frac{\beta^{-\theta}}{2\pi}
B\!\left(\theta+\frac{1}{2},\frac{1}{2}\right)
{}_2F_1\!\left(
\theta,\theta+\frac{1}{2};\theta+1;-\beta^{-1}
\right).
\label{eq:app_HD_Qbar_hypergeom}
\end{equation}
With $B(x,y)=\Gamma(x)\Gamma(y)/\Gamma(x+y)$ and
$\Gamma(1/2)=\sqrt{\pi}$, Eq.~\eqref{eq:app_HD_Qbar_hypergeom} reduces to
\begin{equation}
\bar Q_E(V)
=
\frac{\Gamma\!\left(\theta+\frac{1}{2}\right)}
{2\sqrt{\pi}\,\Gamma(\theta+1)}
\beta^{-\theta}
{}_2F_1\!\left(
\theta,\theta+\frac{1}{2};\theta+1;-\beta^{-1}
\right).
\label{eq:app_HD_Qbar_closed}
\end{equation}

Since $\gamma=x_A+i p_A$, Eq.~\eqref{eq:HD_QBER_reduced} can be expressed as
{\small
\begin{equation}
\mathrm{QBER}_{\mathrm{HD,acc}}
=
\frac{y^m}
{m!\,\pi\,(V_T+1)P_{\mathrm{acc}}}
\int_{-\infty}^{\infty}
\int_{-\infty}^{\infty}
(x_A^2+p_A^2)^m
e^{-w(x_A^2+p_A^2)}
 \times
\bar Q_E\!\left(
\frac{\sqrt{I_p}\mu |x_A|}{\sigma_H}
\right)
dp_A\,dx_A .
\label{eq:app_HD_cartesian}
\end{equation}}
Using the binomial expansion of $(x_A^2+p_A^2)^m$ and the Gaussian
moment integral
\begin{equation}
\int_{-\infty}^{\infty}
p_A^{2r}e^{-w p_A^2}dp_A
=
\frac{\Gamma\!\left(r+\frac{1}{2}\right)}
{w^{r+\frac{1}{2}}},
\label{eq:app_HD_gaussian_moment}
\end{equation}
one obtains
{\small
\begin{align}
\mathrm{QBER}_{\mathrm{HD,acc}}
=
\frac{y^m}
{m!\,\pi\,(V_T+1)P_{\mathrm{acc}}}
\sum_{r=0}^{m}
\binom{m}{r}
\frac{\Gamma\!\left(r+\frac{1}{2}\right)}
{w^{r+\frac{1}{2}}}
\times
\int_{-\infty}^{\infty}
x_A^{2(m-r)}
e^{-w x_A^2}
\bar Q_E\!\left(
\frac{\sqrt{I_p}\mu |x_A|}{\sigma_H}
\right)dx_A .
\label{eq:app_HD_after_p}
\end{align}}
Since the integrand is even in $x$, Eq.~\eqref{eq:app_HD_after_p} reduces to
\begin{equation}
\mathrm{QBER}_{\mathrm{HD,acc}}
=
C_{\mathrm{HD}}
\sum_{r=0}^{m}
\binom{m}{r}
\frac{\Gamma\!\left(r+\frac{1}{2}\right)}
{w^{r+\frac{1}{2}}}
\int_{0}^{\infty}
x_A^{2(m-r)}
e^{-w x_A^2}
\bar Q_E\!\left(
\frac{\sqrt{I_p}\mu x_A}{\sigma_H}
\right)dx_A,
\label{eq:app_HD_final}
\end{equation}
where
\begin{equation*}
C_{\mathrm{HD}}
=
\frac{2y^m}
{m!\,\pi\,(V_T+1)P_{\mathrm{acc}}}.
\label{eq:app_HD_C}
\end{equation*}
This is Eq.~\eqref{eq:HD_QBER_final}.

\bibliography{Biblio}
\end{document}